\documentclass[titlepage, 12pt, a4paper]{article}
\usepackage{amsmath,amssymb}
\usepackage{natbib}
\usepackage{mathtools}
\usepackage{amsthm}

\usepackage{tagging}
\usepackage{array}
\usepackage{appendix}
\usepackage{amssymb}
\usepackage{amsmath}
\usepackage{bm}
\usepackage{graphicx}
\usepackage{amsbsy}
\usepackage[english]{babel}
\usepackage{float}
\usepackage{rotating}
\usepackage{xcolor}
\usepackage{setspace}
\usepackage{soul} 
\usepackage[hidelinks]{hyperref}
\usepackage{tikz}
\usepackage{orcidlink}
\usepackage{algorithm}
\usepackage{AbadirMagnus}

\usepackage[a4paper,top=3cm,bottom=2cm,left=3cm,right=3cm,marginparwidth=2cm]{geometry}

\begin{document}

\title{Severe Testing of Benford's Law}

\author{Roy Cerqueti \orcidlink{0000-0002-1871-7371}\\La Sapienza University of Rome, Rome, Italy\\Department of Social and Economic Sciences\\and London South Bank University Business School, London, UK\\roy.cerqueti@uniroma1.it\\
\and Claudio Lupi\thanks{Corresponding author.} \orcidlink{0000-0001-5166-1130}\\University of Molise, Campobasso, Italy\\Department of Economics\\lupi@unimol.it}

\date{}

\maketitle

\begin{abstract}
Benford's law is often used as a support to critical decisions related to data quality or the presence of data manipulations or even fraud. However, many authors argue that conventional statistical tests will reject the null of  data ``Benford-ness'' if applied in samples of the typical size in this kind of applications, even in the presence of tiny and practically unimportant deviations from Benford's law. Therefore, they suggest using alternative criteria that, however, lack solid statistical foundations. This paper contributes to the debate on the ``large $n$'' (or ``excess power'') problem in the context of Benford's law testing. This issue is discussed in relation with the notion of severity testing for goodness of fit tests, with a specific focus on tests for conformity with Benford's law. To do so, we also derive the asymptotic distribution of the mean absolute deviation ($MAD$) statistic as well as an asymptotic standard normal test. Finally, the severity testing principle is applied to six controversial data sets to assess their ``Benford-ness''.
\\
\\
\textbf{keywords} Benford's law; Data quality; Goodness of fit; Large $n$ problem; Severity.\\
\textbf{MSC} 62E20; 62F03.
\end{abstract}

\onehalfspacing

\section{Introduction}
Benford's law, so named after \cite{benford1938}, predicts that in many real instances the relative frequency of the first significant digits in a set of numbers approaches 
\begin{equation}\label{eq:Benford}
b_i = \Pr(X = d_i) = \log_{10} \left(1 + \frac{1}{d_i} \right)
\end{equation}
with $d_i \in \{1, \ldots , 9 \}$ for the first digit case or $d_i \in \{10, \ldots , 99 \}$ for the first-two digits case.

The reasons why data from many disciplines seem to obey Benford's law has been deeply investigated and some explanations have been put forward. In this respect, \cite{raimi1976}, \cite{hill1995c, hill1995}, \cite{rodriguez2004}, \cite{fewster2009}, \cite{blocksavits2010}, \cite{ross2011}, and \cite{whyman2016} represent important contributions.

Based on this widespread regularity, many critical decisions regarding data quality \citep[see, e.g.,][]{Hand_2018, Kaiser_2019, lietal2019} or the possible presence of data manipulation or fraud, especially in financial data \citep[see, e.g.,][]{chogaines2007, nigrini2012, Ausloos_2017, Barabesi_2018} are based on data (non-)conformity with Benford's law: the idea is that, provided some minimal conditions are met, genuine data should obey Benford's law. Such delicate decisions should be taken using robust and firmly grounded statistical evidence. However, this is not always the case.  

From an empirical point of view, data conformity with Benford's law can be tested using different criteria and statistical tests \citep[see, e.g.,][]{nigrini2012, Joenssen2015, barneyschulzke2016, Barabesi_2018, hassler2019, Barabesi_2021, cerquetilupi2021}. Among the available tests, Pearson's $\chi^2$ test has a predominant role. However, the use of the $\chi^2$ and other goodness-of-fit statistical tests has been forcefully criticized in a large part of the literature on Benford's law on the grounds that these tests have ``excessive power'' --- i.e., given a large enough sample size, they tend to reject the null of conformity with Benford’s law even in the presence of tiny and practically unimportant deviations \citep[see, e.g.,][]{chogaines2007, nigrini2012, tsagbeyetal2017, druicaetal2018, Kossovsky_2021}. For this reason \cite{drake2000computer} and \cite{nigrini2012} favor the use of the Mean Absolute Deviation ($MAD$) criterion, a measure of the average distance between the relative frequencies of the observed digits with Benford's theoretical frequencies:
\begin{equation}
MAD = \frac{1}{k} \sum_{i = 1}^k \left| p_i - b_i \right|
\end{equation}
where $k$ is the number of digits ($k=9$ for the first digit case; $k=90$ for the first-two digits case), $p_i$ are the proportions of the first significant digits in the sample, and $b_i$ are Benford's probabilities from (\ref{eq:Benford}). This quantity (allegedly, see section \ref{sec:severity_benford}) should not depend on the sample size $n$, and should not incur in the ``excess power'' problem. In fact, referring to the $MAD$, \citet[p.~158]{nigrini2012} explicitly argues that 
\begin{quote}
``What is needed is a test that ignores the number of records. The mean absolute deviation ($MAD$) test is such a test [...]''.
\end{quote}
Furthermore, on the basis of personal experience and experimentation on many different data sets, \cite{drake2000computer} and \citet[table~7.1]{nigrini2012} propose some fixed thresholds of the $MAD$ to classify data into four classes of conformity with Benford's law corresponding to ``close conformity'', ``acceptable conformity'', ``marginally acceptable conformity'', and ``nonconformity''. These thresholds have been largely accepted in the applied literature and Nigrini's $MAD$ since then has become a workhorse of Benford's analysis. In our view, basing sensitive decisions that may critically affect different aspects of society on criteria that lack solid statistical foundations is not a satisfactory answer to the large~$n$ (or ``excess power'') problem raised in the literature. In this sense we are very sympathetic with the views expressed, e.g., in \cite{Hand_2021}.

The large~$n$ problem is very well known in the statistical literature \citep[see, e.g.,][to cite a few]{Berkson_1938, Lindley_1957, cohen1994, granger1998} and simply reflects the fact that any consistent test will asymptotically (with $n \to \infty$) reject with probability 1 any fixed false hypothesis, however close to the null this might be. In fact, this should not be considered a ``problem'' at all, but rather a desirable property of the test. Instead, the problem lies in not recognizing that a rejection of the null hypothesis carries a weight of evidence against the null and in favour of a particular alternative that depends on the number of observations $n$.  Given that the power of the test increases with the sample size $n$, a rejection in the presence of a small $n$ (low power) is stronger than one in the presence of a very large number of observations (high power). Giving the same weight to all rejections irrespective of $n$ may bring to the \emph{fallacy of rejection} in the sense of \cite{mayospanos2006, MayoSpanos_2011}. What is of interest to us is the possibility of assessing the largest discrepancy from the null hypothesis warranted by the data at hand: then, depending on the specific context, we will be able to evaluate if that discrepancy is substantive or not.

In some circumstances we might be interested in rejecting the null (in favour of a specific alternative) only in the presence of sizeable deviations from the null: this is certainly true in the case of tests of conformity with Benford's law. The solution we propose is at the same time rigorous, elegant, and simple: we propose that the \emph{severity principle} \citep[see, e.g.,][]{mayospanos2006, mayospanos2010, MayoSpanos_2011, Mayo2018} should be routinely applied when testing for goodness-of-fit in general, and for conformity with Benford's distribution in particular. Severity allows us to focus on the presence (or absence) of what we may regard as a \emph{substantive} discrepancy. In Spanos's words

\begin{quote}
``The severity evaluation is a post-data appraisal of the accept/reject and $p$ value results with a view to provide an evidential interpretation'' \citep[][p.~647]{spanos2014}.
\end{quote}

In this paper we show how the post-data severity evaluation (PDSE) can be conveniently implemented in the special case of tests of conformity with Benford's law. In our view, the severity principle finds a perfect application in goodness-of-fit testing in general, and in the assessment of Benford's law in particular.

The structure of the paper is as follows: the next section illustrates the severity principle. Section~\ref{sec:severity_benford} describes how the severity principle can be applied to the case of tests of Benford's law. In section~\ref{sec:application} we apply the severity principle in testing Benford's law on some controversial dataset \citep{Kossovsky_2021} to shed some light on their ``Benford-ness''. Section~\ref{sec:conclusions} draws some conclusions.


\section{The severity principle}\label{sec:severity}
Despite having been introduced in the 1990s \citep[see, e.g.,][]{mayo1996} and being potentially important in many real-world applications, the severity principle is still unfamiliar to most statisticians. The knowledge of this important concept has been relegated mostly among those interested in the relationships between statistics and the philosophy of science. Unsurprisingly (but regretfully), papers adopting the severe testing approach are virtually absent in the applied statistical literature.

Rejecting the null hypothesis on the basis of a small (or even very small) test $p$~value is not informative of the size of the discrepancy existing between the null and the data at hand. Very powerful tests (e.g. in the presence of a very large sample) may reject the null even for practically unimportant deviations. The idea underlying the concept of severity is that rejection (or non-rejection) \emph{per se} is not sufficient to severely test the null hypothesis.

Given the delicate decisions that can be taken on the basis of (non-)conformity with Benford's law and the typically large sample sizes used to test Benford's law, here we are mostly concerned with what \cite{mayospanos2006} call the \emph{fallacy of rejection}:

\begin{quote}
``Evidence \emph{against} $H_0$ is misinterpreted as evidence \emph{for} a particular $H_1$. This fallacy arises in cases where the test in question has high power to detect substantively minor discrepancies. Since the power of a test increases with the sample size $n$, this renders N-P [Neyman-Pearson] rejections, as well as tiny $p$~values, with large $n$, highly susceptible to this fallacy.'' \citep[p.~599]{spanos2019}
\end{quote}

The idea of severity testing revolves around the intuition that rejection of the null hypothesis based on a test with low power for detecting a discrepancy $\gamma$ provides stronger evidence for the presence of a discrepancy $\gamma$ than rejection based on a substantially more powerful test. By its very nature, severity is based on a post-data perspective, when the direction of the possible discrepancy is known.

We are interested in the severity with which claim $C$ passes test $T$ with an outcome $T(\vx)$ \citep{mayospanos2006, Mayo2018}:
\begin{equation}
SEV\left[ \mbox{test } T, \mbox{outcome } T(\vx), \mbox{claim } C \right]
\end{equation}
which will be abbreviated simply as $SEV[T(\vx)]$ for ease of notation.
For the claim $C$ to pass a severe test with data $\vx$ we require that
\begin{enumerate}
\item data $\vx$ agree with claim $C$, and
\item \emph{with very high probability}, test $T$ would have produced a result that accords \emph{less} with claim $C$ if $C$ is false.
\end{enumerate}

To explain the severity principle in relation with the sample size, we start from an example borrowed and adapted from \cite{mayospanos2010}. 

Assume that we are interested in a one-sided test of the mean. The sample $\mX := (X_1, \ldots, X_n)$ is such that $X_i \distr NIID(\mu, \sigma^2)$ with known $\sigma^2 = 4$ (for simplicity). The test statistic is $T_n(\mX) := (\bar X - \mu_0) / \sigma_{\bar X}$ where $\bar X$ is the sample mean and $\sigma_{\bar X} = \sigma / \sqrt{n}$. The null and the alternative hypothesis of interest are $\rH_0 : \mu = \mu_0 = 0$ and $\rH_1 : \mu = \mu_1 = \mu_0 + \gamma > 0$, respectively. Under the null, $T_n(\mX) \distr \rN(0, 1)$.

Suppose that we test the null hypothesis on four observed samples ($\vx_1$, $\vx_2$, $\vx_3$, and $\vx_4$) of different size ($n_1 = 100$, $n_2 = 200$, $n_3 = 500$, and $n_4 = 1000$). Suppose further that in each sample we obtain an identical result, $T_{n_i}(\vx_i) = 2$ $\forall i \in \{1, \ldots, 4\}$. Using the usual significance levels, the four tests point to the rejection of the null hypothesis, with a $p$~value equal to $0.0228$. However, this result is not informative about the size of the possible discrepancy with the null.

Assume that we are interested in the existence of a \emph{substantive} discrepancy $\gamma$ that, in our specific context, we suppose that can be quantified as $\gamma > 0.2$. How robust is the inference that $\mu = \mu_0 + \gamma > 0.2$ in the four samples, given that the test statistic in each sample is equal to 2? We have to consider the severity of the claim $C: \mu > 0.2$, with a test outcome 2 using our test $T$. The fact that the test rejects the null is consistent with claim $C$. However, we need also that with high probability the test outcome should agree less with $C$ than the actual outcome does, if $C$ is false. This means that we must evaluate
\begin{equation}\label{eq:severity1}
SEV[T_n(\mX)] = \Pr \left[T_n(\mX) \leq T_n(\vx_o); \mu > \mu_1 \mbox{ is false}\right]
\end{equation}
where $T_n(\vx_o)$ is the test statistic calculated on the observed sample and ``;'' means ``calculated assuming'' the specified condition. Of course, saying that $\mu > \mu_1$ is false is equivalent to saying that $\mu \leq \mu_1$ is true. In fact, it is sufficient to compute
\begin{equation}
SEV[T_n(\mX)] = \Pr \left[T_n(\mX) \leq T_n(\vx_o); \mu = \mu_1 \right]
\end{equation}
because for any $\mu < \mu_1$ we would obtain larger values of the severity.

\begin{figure}
\centering
\includegraphics[scale = 0.37]{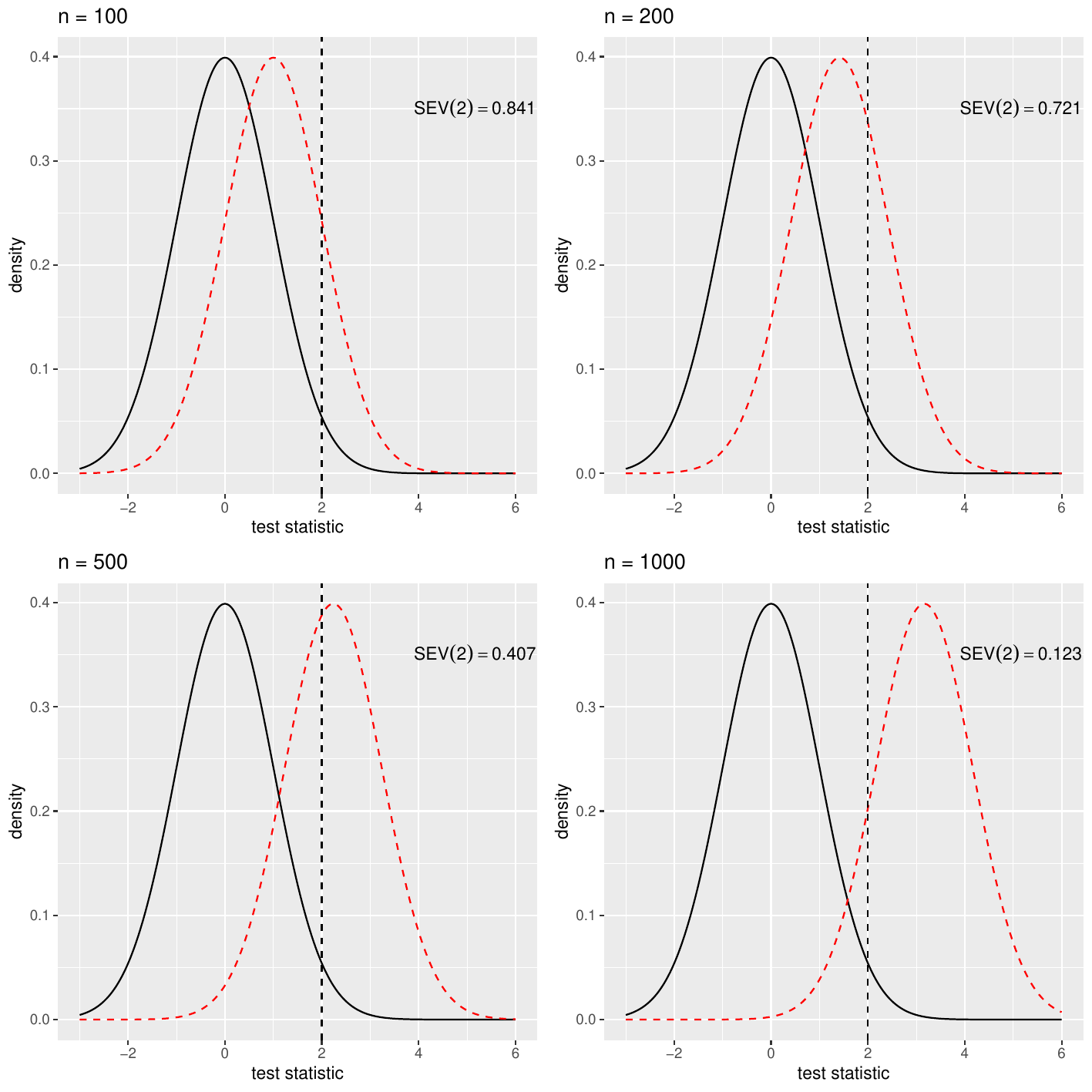}
\vspace{-0.1cm}\caption{Severity of the test in four samples of different size. The solid curve is the distribution of the test statistic under the null $\mu = \mu_0 = 0$; the dashed curve under the specific alternative $\mu = \mu_0 + \gamma = 0.2$. The vertical dashed line represents the value of the test statistic. The reported values of $SEV(2)$ represent the severity of the test in the different samples.\label{fig:severity_example}}
\end{figure}

The situation is summarized in figure~\ref{fig:severity_example}. When $\mu = \mu_1 = \mu_0 + \gamma$ is assumed true, the distribution of $T_n(\mX)$ becomes $\rN[\sqrt{n}\,\gamma/\sigma, 1]$. The effect is that the distribution shifts to the right for higher values of $n$. Therefore, the rejection of the null implies different levels of severity (with respect to our alternative of interest) for different sample sizes: severity is higher for smaller values of $n$, ranging from 0.841 ($n = 100$) to 0.123 ($n = 1000$). That is, despite the fact that the null hypothesis is rejected at the same significance level, evidence in favour of our alternative of interest becomes weaker as $n$ grows. 

\begin{figure}
\centering
\includegraphics[scale = 0.5]{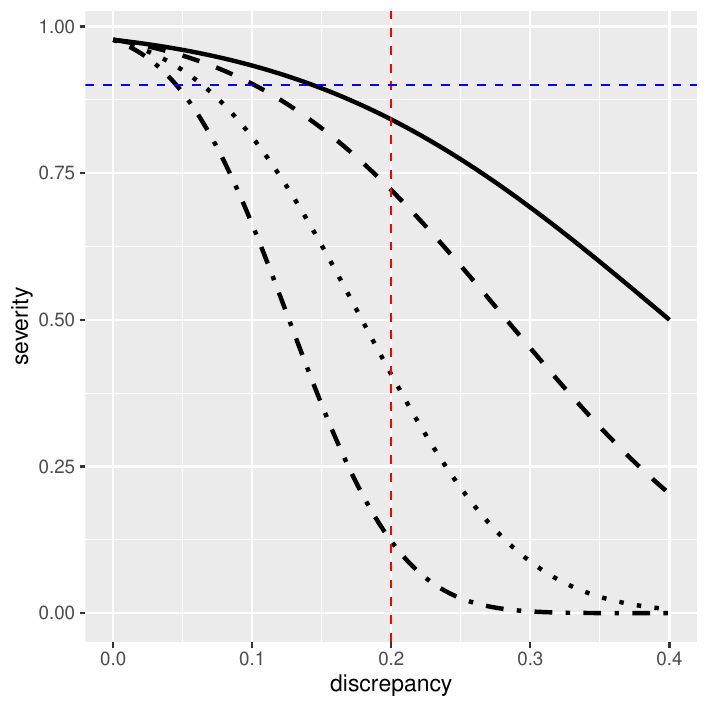}
\vspace{-0.1cm}\caption{Severity of the test as a function of discrepancy $\gamma$ in four samples of different size. The curves represent $\Pr \left[T_n(\mX) \leq T_n(\vx_o) \right]$ with $T_n(\mX) \distr \rN[\sqrt{n} \, \gamma/\sigma, 1]$ with $n = 100$ (solid), $n = 200$ (dashed), $n = 500$ (dotted), $n = 1000$ (dot-dashed). The dashed vertical line represent the hypothesized discrepancy of interest, $\gamma = 0.2$. The dashed horizontal line corresponds to a severity level equal to 0.9. \label{fig:severity_function_n}}
\end{figure}

Put under another perspective, the combined effect of discrepancy and sample size on severity can be visualized as in figure~\ref{fig:severity_function_n}, where severity is plotted as a function of the discrepancy $\gamma$ for four different values of the sample size $n$. From figure~\ref{fig:severity_function_n} it is easy to see that the alternative hypothesis $H_1: \gamma > 0.2$ is supported with decreasing severity as $n$ increases. The figure also highlights that different discrepancies can be supported with the same severity for different values of $n$. For example, for $n = 100$, the null $H_0: \gamma = 0$ can be rejected with severity 0.9 in favour of the alternative $H_1: \gamma > 0.144$, whereas in the presence of $n = 1000$ the null can be rejected with the same severity in favour of the alternative $H_1: \gamma > 0.045$. Read in this way, figure~\ref{fig:severity_function_n} allows us to evaluate severity of rejection in a flexible way, leaving to the expert's judgement the assessment of the practical importance of the rejection. In so doing, we are in line with the expert judgement approach for identifying the thresholds of close/acceptable/marginally acceptable conformity with Benford's Law proposed by \cite{nigrini2012} for the case of $MAD$ and \cite{kossovsky2014benford} for the case of the sum of squared deviations criterion \citep[see also][for additional discussion on this point]{Cerqueti_2021}.


\section{Severe tests of Benford's law}\label{sec:severity_benford}
\subsection{Normal test}
In order to maintain a direct relation with Nigrini's $MAD$, it would be interesting to use a $MAD$-based test in evaluating data ``Benford-ness''.

\cite{cerquetilupi2021} show that, in the presence of a random sample, under the null hypothesis of data conformity with Benford's law
\begin{eqnarray} \label{eq:single}
\sqrt{n}\frac{\left| p_i - b_i \right|}{\sqrt{b_i (1 - b_i)}} &=& \sqrt{n}\frac{\left| e_i \right|}{\sqrt{b_i (1 - b_i)}} \nonumber \\
&\dto& \rN \left(\sqrt{\frac{2}{\pi}}, 1 - \frac{2}{\pi}  \right) 
\end{eqnarray}
as $n \to \infty$.

Equation~(\ref{eq:single}) can be written in vector form as
\begin{equation}\label{eq:vector}
\sqrt{n} \mD^{-1} | \ve | \dto \rN \left(\sqrt{\frac{2}{\pi}} \vones, \mR  \right)
\end{equation}
where $\mD = \diag \left(\sqrt{b_i (1 - b_i)}\right)$ (with $i = 1, \ldots, k$), $\ve = (e_1, \ldots, e_k)^\prime$, $\vones$ is a $k$-vector of 1s, and the covariance matrix $\mR$ is defined as \citep[see][]{cerquetilupi2021}
\begin{equation}
\mR = \left(
\begin{array}{cccc}
r_{11} & r_{12} & \ldots & r_{1k} \\ 
r_{12} & r_{22} & \ldots & r_{2k} \\ 
\vdots & \vdots & \ddots & \vdots \\ 
r_{1k} & r_{2k} & \ldots & r_{kk}
\end{array} 
\right) = \left\lbrace r_{ij} \right\rbrace
\end{equation}
with 
\begin{equation}
r_{ij} = \frac{2}{\pi} \left( \rho_{ij} \arcsin(\rho_{ij}) + \sqrt{1 - \rho_{ij}^2} \right) - \frac{2}{\pi} 
\end{equation} 
where
\begin{equation}
\rho_{ij} = \left\lbrace \begin{array}{l}
- \sqrt{\frac{b_i b_j}{(1 - b_i)(1 - b_j)}} \;\; \mbox{ for } i \neq j \\ 
1 \;\; \mbox{ for } i = j
\end{array} \right.
\end{equation}

It follows that
\begin{equation}
\sqrt{n} \left| \ve \right| \dto \rN \left(\sqrt{\frac{2}{\pi}} \mD \vones, \mD \mR \mD  \right)
\end{equation}
and
\begin{equation}\label{eq:asymptotic_distr}
\sqrt{n} MAD = \frac{\sqrt{n}}{k} \vones^\prime \left| \ve \right| \dto \rN \left(\sqrt{\frac{2}{\pi k^2}} \vones^\prime \mD \vones, \frac{1}{k^2} \vones^\prime \mD \mR \mD  \vones \right) \,.
\end{equation}

Equation~(\ref{eq:asymptotic_distr}) represents the asymptotic distribution of the $MAD$ under the null hypothesis that Benford's law is valid. Finally, if Benford's law holds, for $n$ large Nigrini's $MAD$ is approximately distributed as 
\begin{equation}\label{eq:approximate_distr}
\rN \left(\sqrt{\frac{2}{\pi n k^2}} \vones^\prime \mD \vones, \frac{1}{n k^2} \vones^\prime \mD \mR \mD  \vones \right) \,.
\end{equation}

Therefore, it is clear that Nigrini's $MAD$, far from being independent of $n$, under the null hypothesis of conformity with Benford's law is such that its expected value and its standard deviation are inverse functions of $\sqrt{n}$. To highlight that both the $MAD$ and its expected value depend on $n$ and $k$, from now on we will denote them as $MAD_{n,k}$ and $E(MAD_{n,k})$, respectively.

The use of the fixed thresholds proposed in \citet[][table~7.1]{nigrini2012} implies that the $MAD_{n,k}$ criterion will tend to reject too often for small values of $n$ and to be too conservative for large ones.

Following \cite{barneyschulzke2016}, our approach is to measure discrepancy in terms of the excess $MAD$
\begin{equation}
\delta_{n,k} = MAD_{n,k} - E(MAD_{n,k}) \,.
\end{equation}

Note that by~(\ref{eq:asymptotic_distr}), under the null hypothesis of data conformity with Benford's law we have
\begin{eqnarray}\label{eq:testStat}
\tilde \delta &=& \frac{k \sqrt{n} \, \delta_{n,k}}{{\sqrt{\vones^\prime \mD \mR \mD  \vones}}} \nonumber \\
&=& \frac{k \sqrt{n} \left[ MAD_{n,k} - \sqrt{\frac{2}{\pi k^2}} \vones^\prime \mD \vones \right]}{\sqrt{\vones^\prime \mD \mR \mD  \vones}} \dto \rN(0,1) \, .
\end{eqnarray}

In this way we can obtain a standard normal test of conformity based on a scaled version of the excess $MAD$. We call this test ``excess $MAD$ test''. The null of conformity is $\rH_0: \delta_{n,k} = 0$; the alternative of interest is one-sided, $\rH_1: \delta_{n,k} > 0$ and can be specialized as $\rH_1: \delta_{n,k} > \delta^* > 0$ to take into account substantive discrepancy. In fact, it should be stressed that the test $p$~value is a post-data error probability for which we know the relevant tail so that we can legitimately disregard cases for which the test statistic is negative. In the present case, the distribution of the test statistic (\ref{eq:testStat}) under the alternative $H_1: \delta_{n,k} = \delta^*$ is approximately $\rN \left(k \sqrt{n} \, \delta^* / \sqrt{\vones^\prime \mD \mR \mD \vones}, 1 \right)$. The distribution of the test statistic under the alternative can be used to carry out a post-data severity evaluation as in section~\ref{sec:severity}.


\subsection{Chi-square test}
Pearson's $\chi^2$ goodness-of-fit test
\begin{equation}\label{eq:chi-square}
Q_n := n \sum_{i=1}^k \frac {\left( p_{i} - b_{i} \right)^2}{b_i}
\end{equation}
is one of the most widely used tests to check data ``Benford-ness''. Under the null that data conform to Benford's law (i.e., $p_i = b_i$, $\forall i = 1, \ldots, k$), the test statistic is asymptotically distributed as $\chi^2(k-1)$, a central chi-square random variable with $k-1$ degrees of freedom, where $k = 9$ or $k = 90$ in the first digit or the the first-two digits case, respectively. Under the alternative $p_i \neq b_i$. More precisely, assume that
\begin{eqnarray}
p_i &=& b_i + \epsilon_i \nonumber \\
\epsilon_i &=& n^{-\,\frac{1}{2}} h_i \nonumber
\end{eqnarray}
with $\sum_{i=1}^k \epsilon_i = 0$: then for large $n$ the test statistic is approximately distributed as a non-central $\chi^2(k-1, \psi)$ random variable, where the non-centrality parameter $\psi$ is \citep[see, e.g.,][]{lehmannromano2005}
\begin{equation}\label{eq:noncentrality}
\psi = n \sum_{i = 1}^k \frac{\epsilon_i^2}{b_i} \,. 
\end{equation}
The post-data severity evaluation exemplified in section~\ref{sec:severity} for the normal case can be used also in the presence of a chi-square statistical test, using the appropriate distributions. Again, it's up to the expert's judgement the definition of what constitutes a practically relevant discrepancy from the null.


\section{Application}\label{sec:application}

In a recent paper, \citet[][p.~426]{Kossovsky_2021} argues that there are
\begin{quotation}
``[...] six data sets wrongly rejected as non-Benford by the chi-square test:
\begin{enumerate}
\item Time in seconds between the 19,451 Global Earthquakes during 2012;
\item USA Cities and Towns in 2009 --- 19,509 Population Centers;
\item Canford PLC Price List, Oct 2013 --- 15,194 items for sale;
\item 48,111 Star Distances from the Solar System, NASA;
\item Biological Genetic Measure--DNA V230 Bone Marrow --- 91,223 data points;
\item Oklahoma State 986,962 positive expenses below \$1 million in 2011.
\end{enumerate}
The above six data sets are truly excellent real-life Benford examples, where the digital phenomenon manifests itself decisively, for the 1st digit distribution as well as for the 2nd digit distribution, yet, the chi-square test claims that they are not complying with the law of Benford. [...] These six data sets are excellent representatives of the phenomenon, with deep loyalty to Benford's Law, yet all are being betrayed by the chi-square test.''
\end{quotation}

In this section we apply the severity principle to the excess MAD test (\ref{eq:testStat}) and to the chi-square test (\ref{eq:chi-square}) to verify the possible existence of substantive discrepancies between the data and Benford's law. Data are courtesy of Alex Kossovsky, and have been downloaded from \url{https://web.williams.edu/Mathematics/sjmiller/public_html/benfordresources/}.


\subsection{Excess $MAD$ test}
Table~\ref{tab:excessMAD} reports the results of the excess $MAD$ normal test (\ref{eq:testStat}) for each of the six data sets under examination for both the first digit and the first-two digits case. Most of the tests point clearly towards the rejection of the null of conformity with Benford's law, as highlighted by \cite{Kossovsky_2021} with reference to the chi-square test, with the only exception of the first-two digits test applied to the USA city population data set. However, the values of the $MAD_{n,k}$ criterion are always well below the non-conformity threshold, i.e. 0.015 for the first digit case and 0.0022 for the first-two digits case \citep[][p.~160]{nigrini2012}. In fact, \cite{Kossovsky_2021} argues that rejections with these data sets are just the manifestation of the ``excess power'' problem.

\begin{table*}
\caption{Excess $MAD$ tests of conformity with Benford's law for Kossovsky's data.}
\label{tab:excessMAD}
\resizebox{1.0 \textwidth}{!} {
\begin{tabular}{|l|r|rrrr|rrrr|}
\hline
                       &   $n$ & \multicolumn{4}{|c|}{first digit} & \multicolumn{4}{c|}{first-two digits} \\
Description            &   & $MAD_{n,9}$ &$\delta_{n,9}$&$\tilde \delta$& $p$~value& $MAD_{n,90}$ &$\delta_{n,90}$&$\tilde \delta$& $p$~value \\
\hline
Earthquakes            &   19,451 & 0.00479 & 0.00311 &  6.621 &   0.00000 & 0.00080 & 0.00023 &   4.873 &  0.00000 \\
USA city population    &   19,509 & 0.00312 & 0.00144 &  3.065 &   0.00109 & 0.00062 & 0.00005 &   1.018 &  0.15424 \\
Canford PLC Price List &   15,194 & 0.00517 & 0.00327 &  6.146 &   0.00000 & 0.00149 & 0.00085 &  15.591 &  0.00000 \\
Star distances         &   48,111 & 0.01088 & 0.00981 & 32.839 &   0.00000 & 0.00125 & 0.00089 &  29.267 &  0.00000 \\
Genetic measure        &   91,223 & 0.00187 & 0.00109 &  5.034 &   0.00000 & 0.00035 & 0.00008 &   3.765 &  0.00008 \\
Oklahoma State expenses&  967,152 & 0.00253 & 0.00229 & 34.394 &   0.00000 & 0.00208 & 0.00200 & 293.700 &  0.00000 \\
\hline
\end{tabular}
}
\end{table*}

\begin{figure}
\centering
\includegraphics[scale=0.37]{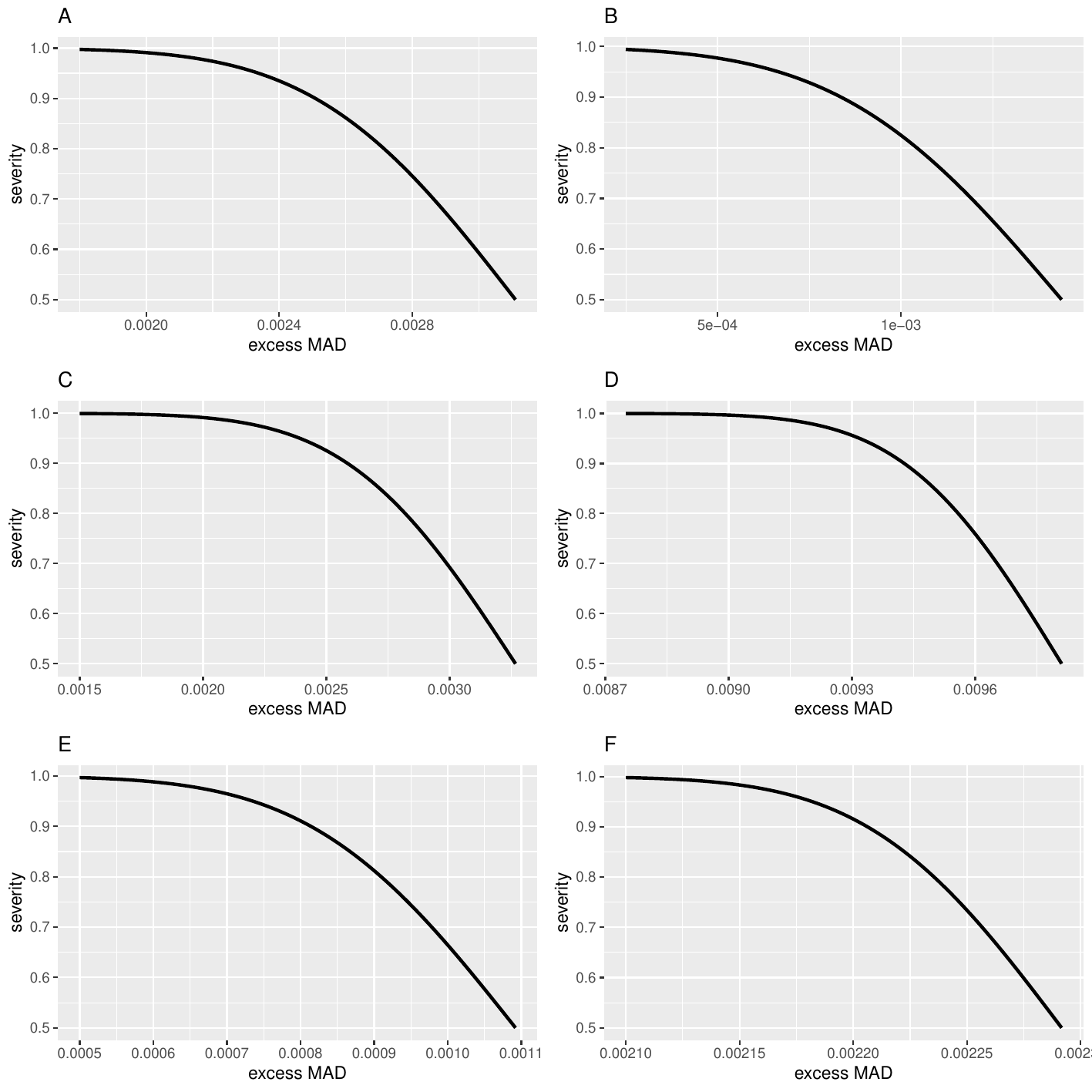}
\vspace{-0.2cm}\caption{Severity of the excess MAD test on the first digit for Kossovsky's data. A, Earthquakes; B, USA city population; C, Canford PLC Price List; D, Star distances; E, Genetic measure; F, Oklahoma State expenses. \label{fig:severity_excessMAD_1}}
\end{figure}

\begin{figure}
\centering
\includegraphics[scale=0.37]{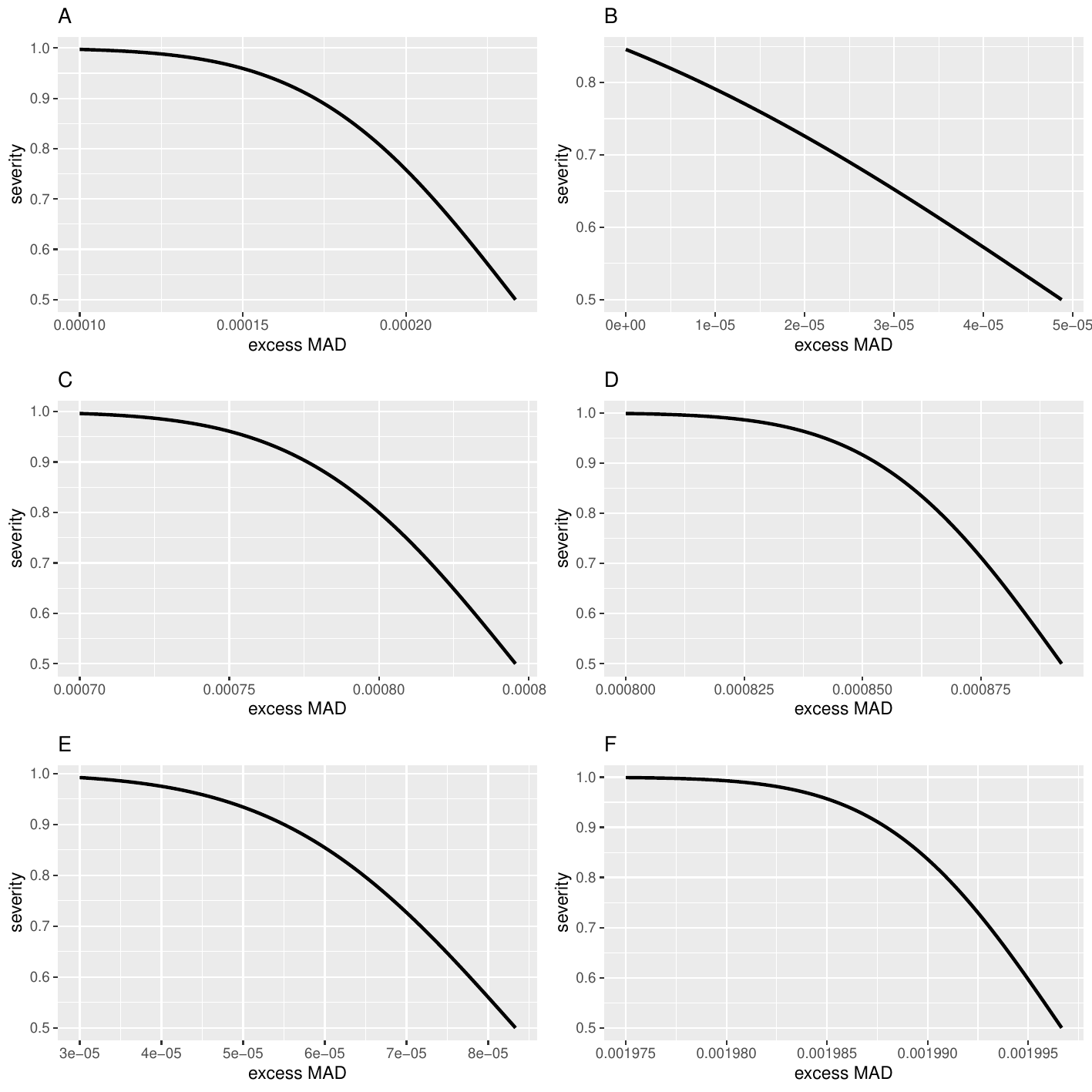}
\vspace{-0.2cm}\caption{Severity of the excess $MAD$ test on the first-two digits for Kossovsky's data. A, Earthquakes; B, USA city population; C, Canford PLC Price List; D, Star distances; E, Genetic measure; F, Oklahoma State expenses. \label{fig:severity_excessMAD_12}}
\end{figure}

Figures~\ref{fig:severity_excessMAD_1} and~\ref{fig:severity_excessMAD_12} report the severity level of the tests on the six data sets for varying values of the excess $MAD$ for the first digit and the first-two digits case, respectively. In order to decide if the null hypothesis is severely rejected or not, we have to refer to some specific value of the excess $MAD$ that we may consider as evidence of a substantive deviation from Benford's distribution. The exact value of this threshold should be left to the expert's judgement. However, without necessarily endorsing Nigrini's thresholds, here we note that the average distance between Nigrini's fixed thresholds and the expected $MAD$ for $n$ varying over a large set of values is
\begin{eqnarray}\label{eq:deltaStar}
\delta_k^* &=& \frac{1}{n_{\max} - n_{\min_k} + 1} \sum_{n = n_{\min_k}}^{n_{\max}} \left(t_k - E(MAD_{n,k}) \right) \nonumber \\
&\approx& \left\lbrace \begin{array}{l}
0.01 \mbox{ in the first digit case}\\ 
0.0012 \mbox{ in the first-two digits case}
\end{array}  \right.
\end{eqnarray}
where $t_k$ $(k \in \{9, 90\})$ is Nigrini's threshold, $E(MAD_{n,k})$ is the expected value of the $MAD$ for each specific number of observations ($n$) and digits ($k$), $n_{\max} = 50000$, and $n_{\min}$ is such to guarantee a minimum expected number of observations in each cell equal to 5 (i.e., $n_{\min} = 110$ in the first digit case and $n_{\min} = 1146$ in the first-two digits case). $t_k$ are taken as Nigrini's thresholds for marginally acceptable conformity, i.e. 0.012 and 0.0018 for the first digit and the first-two digits, respectively. We will conventionally use the $\delta^*$ values resulting from (\ref{eq:deltaStar}) as the reference values to assess the existence of substantial departures from Benford's law in our empirical exercise.

\begin{figure}
\centering
\includegraphics[scale=0.28]{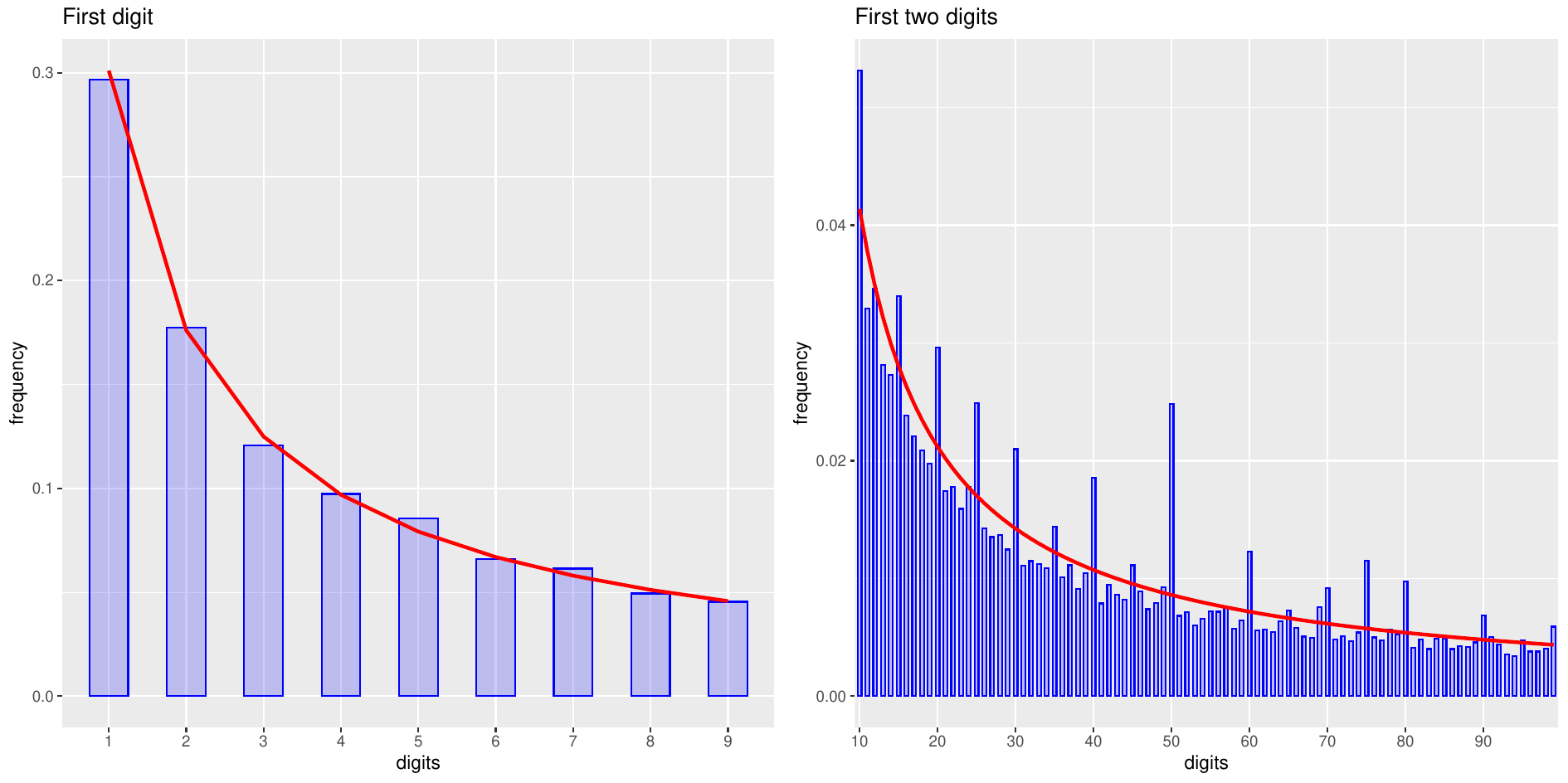}
\vspace{-0.2cm}\caption{Oklahoma State expenses data: first digit (left) and first-two digits (right) distribution. The bars represent the empirical distributions, the solid curves symbolize Benford's distributions.\label{fig:expenses}}
\end{figure}

With reference to the first-digit case, looking at table~\ref{tab:excessMAD} and figure~\ref{fig:severity_excessMAD_1}, we note that, despite the fact that most of the $p$~values of the excess MAD test are tiny, nevertheless  the largest discrepancy warranted by the data with high severity is always well below the ``substantial discrepancy threshold'' $\delta_9^*$. With reference to the first-two digits case, we find a discrepancy larger than $\delta_{90}^*$ with high severity only for the Oklahoma State expenses data set (figure~\ref{fig:severity_excessMAD_12}, panel~F). A closer examination of this data set reveals that there is a clear tendency towards ``round'' numbers: 10, 15, 20, 25, and so forth: see figure~\ref{fig:expenses}. In this case, Nigrini's $MAD$ ($MAD_{n,90} = 0.00208$) suggests ``marginally acceptable conformity'': we believe that this conclusion is totally misleading.

\subsection{Chi-square test}

We repeat the analysis of the six data sets using the standard chi-square test (\ref{eq:chi-square}). We again use Nigrini's non-conformity thresholds as reference values. Simulating a large number of ``perturbed'' Benford's distributions such that the $MAD$ is equal to Nigrini's threshold, it is possible to compute the corresponding average non-centrality parameter of the chi-square distribution. The non-centrality parameters turn to be $0.0238 \times n$ and $0.0474 \times n$ for the first digit and the first-two digits case, respectively. The test statistics and the $p$~values are reported in table~\ref{tab:chisquare}, along with the values of the $MAD$ criterion for ease of comparison. Again, the null is rejected for all the data sets, with the exception of the first-two digits distribution of the USA city population data set.

\begin{table*}
\caption{Chi-square tests of conformity with Benford's law for Kossovsky's data.}
\label{tab:chisquare}
\resizebox{1.0 \textwidth}{!} {
\begin{tabular}{|l|r|rrr|rrr|}
\hline
                       &   $n$ & \multicolumn{3}{|c|}{first digit} & \multicolumn{3}{c|}{first-two digits} \\
Description            &          &   $MAD_{n,9}$ &chi-square& $p$~value&   $MAD_{n,90}$ &chi-square& $p$~value \\
\hline
Earthquakes            &   19,451 & 0.00479 &  53.003 &   0.00000 & 0.00080 &   146.066 &  0.00013 \\
USA city population    &   19,509 & 0.00312 &  17.524 &   0.02510 & 0.00062 &   108.635 &  0.07712 \\
Canford PLC Price List &   15,194 & 0.00517 &  57.739 &   0.00000 & 0.00149 &   487.609 &  0.00000 \\
Star distances         &   48,111 & 0.01088 & 538.107 &   0.00000 & 0.00125 &   862.868 &  0.00000 \\
Genetic measure        &   91,223 & 0.00187 &  34.686 &   0.00003 & 0.00035 &   145.962 &  0.00013 \\
Oklahoma State expenses&  967,152 & 0.00253 & 955.709 &   0.00000 & 0.00208 & 79525.540 &  0.00000 \\
\hline
\end{tabular}
}
\end{table*}

\begin{figure}
\centering
\includegraphics[scale=0.37]{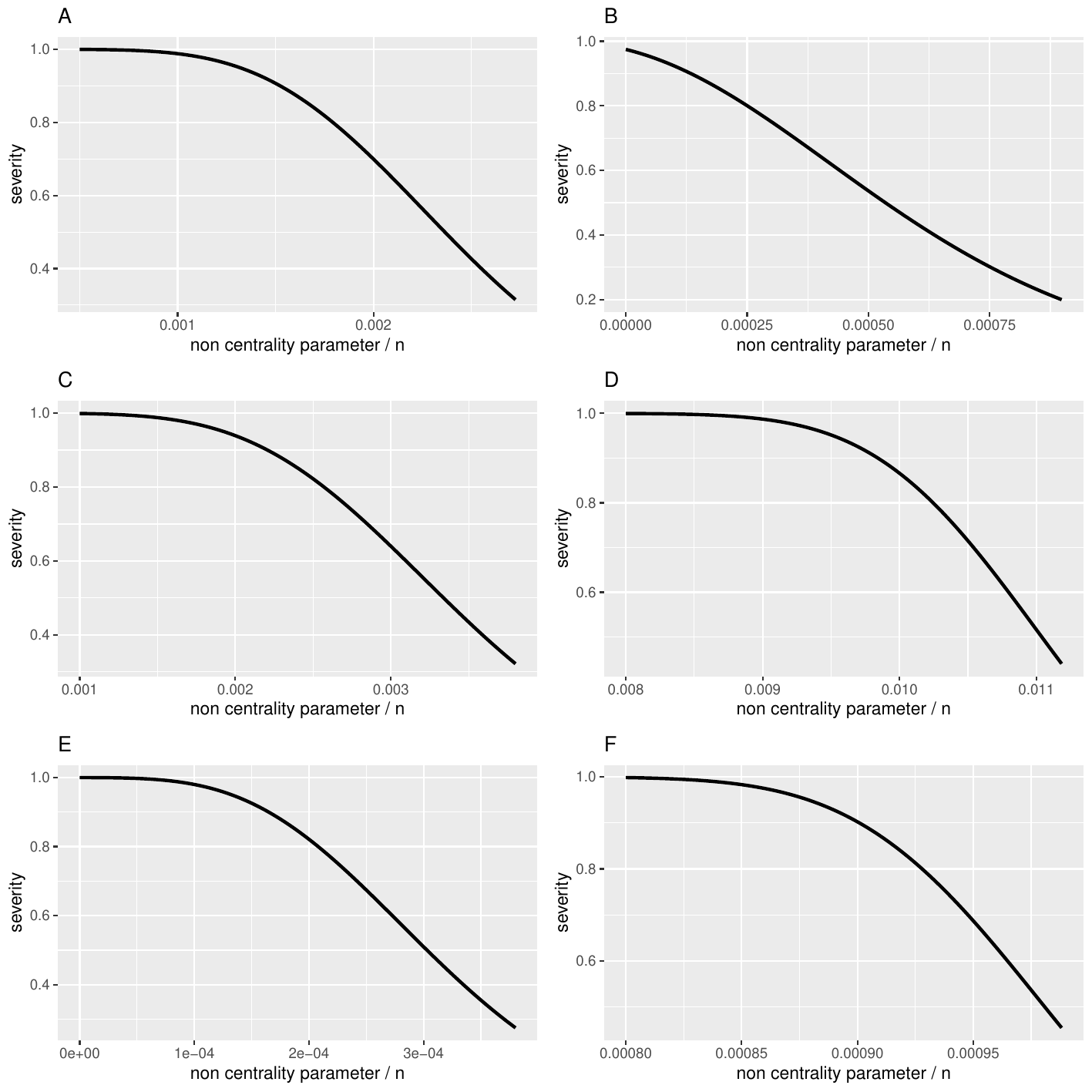}
\vspace{-0.2cm}\caption{Severity of the chi-square test on the first digit for Kossovsky's data. A, Earthquakes; B, USA city population; C, Canford PLC Price List; D, Star distances; E, Genetic measure; F, Oklahoma State expenses. \label{fig:severity_chisquare_1}}
\end{figure}

\begin{figure}
\centering
\includegraphics[scale=0.37]{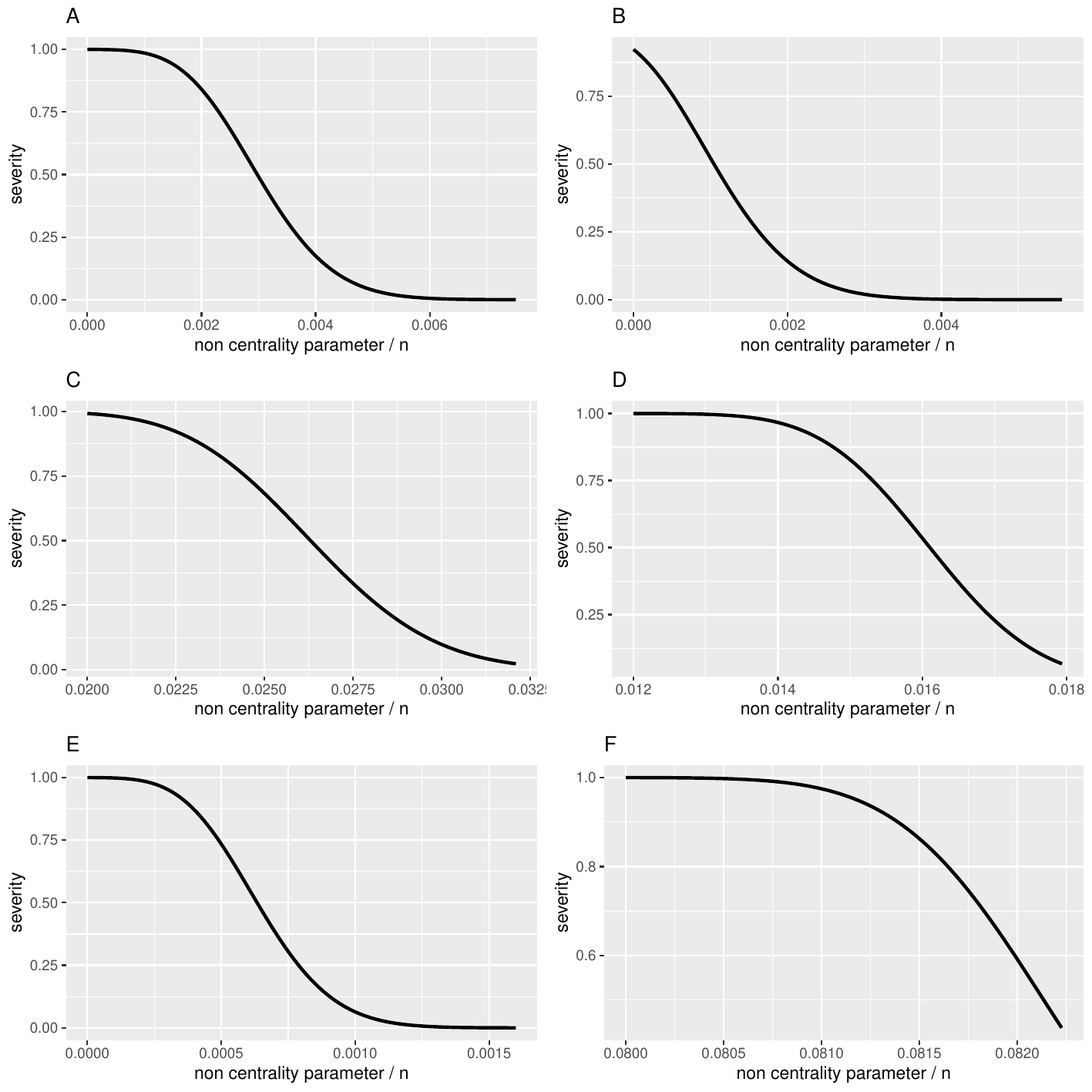}
\vspace{-0.2cm}\caption{Severity of the chi-square test on the first-two digits for Kossovsky's data. A, Earthquakes; B, USA city population; C, Canford PLC Price List; D, Star distances; E, Genetic measure; F, Oklahoma State expenses. \label{fig:severity_chisquare_12}}
\end{figure}

Figures~\ref{fig:severity_chisquare_1} and~\ref{fig:severity_chisquare_12} plot severity of the chi-square tests against the non centrality parameter: to make plots more directly comparable, the non centrality parameter is divided by $n$. As in the case of the excess $MAD$ test, severity is high for fairly large discrepancies only in the case of the first-two digits test for the Oklahoma State expenses data sets (figure~\ref{fig:severity_chisquare_12}, panel~F).


\section{Concluding remarks}\label{sec:conclusions}
Benford's law \citep[after][]{benford1938} is used in practice to support critical decisions in several different contexts, including assessing the possible existence of data manipulation or fraud. These decisions must rely on well founded and trustworthy tests of data conformity with Benford's law. However, many authors \citep[see, e.g.,][]{chogaines2007, nigrini2012, tsagbeyetal2017, druicaetal2018, Kossovsky_2021} have forcefully argued against using statistical tests to assess data conformity with Benford's distribution, because of their ``excess power'' in the presence of large data sets. Alternative decision criteria, such as the mean absolute deviation \citep[$MAD$: see, e.g.,][]{drake2000computer, nigrini2012}, have been proposed in the literature; however, they seem to lack firm statistical foundations.

This paper addresses the ``excess power'' (or ``large $n$'') controversy in the literature related to Benford's law testing. We show that the post-data severity evaluation \citep{mayospanos2006, mayospanos2010, MayoSpanos_2011, Mayo2018} can and should be used to assess data ``Benford-ness'' in large data samples. In order to do so, we also derive the asymptotic distribution of Nigrini's mean absolute deviation (MAD) statistic and propose an asymptotically standard normal test to which the severity principle can be easily applied, even using the information embodied in Nigrini's judgmental thresholds. Finally, we carry out severe testing of Benford's law on six controversial data sets \citep[see][]{Kossovsky_2021} using both the newly proposed normal test as well Pearson's chi-square test. With the exception of the Oklahoma State expenses data set on which we disagree, we otherwise concur with \cite{Kossovsky_2021} that the discrepancies with Benford's law in the majority of these data sets are not substantial. Rejections of the null hypothesis clearly indicate the presence of \emph{some} discrepancies, but their size may be regarded as practically unimportant.

\section*{Acknowledgements}
We would like to express our gratitude to Aris Spanos for his comments and suggestions on an early draft of this paper. Comments from Marcel Ausloos are gratefully acknowledged. All computations have been carried out using R 4.1.2 \citep{R}: graphs greatly benefited from package ``ggplot2'' \citep{ggplot2}. We are grateful to the authors of the software for having provided such excellent tools. We owe a special thank to Alex Kossovsky for having made public his data.
%

\end{document}